\documentclass[twocolumn,prl,nofootinbib]{revtex4}
\usepackage{latexsym}
\usepackage{dcolumn}
\usepackage{epsfig}
\RequirePackage{graphicx}
\def\bea{\begin{eqnarray}} 
\def\eea{\end{eqnarray}} 
\def\rd{{\mathrm d}} 
 
\def\intd4x{\int{\rd}^4x}

\def\m32{{m_{3/2}}}

\newcommand{\la}{\raisebox{-.8ex}{\,$\stackrel{\textstyle <}{\sim}$}\,}

\def\lpa{\lambda_{p{-}\rm air}}
\def\spa{\sigma_{p{-}\rm air}}
\def\spai{\sigma_{p{-}\rm air}^{\rm inel}}
\def\spae{\sigma_{p{-}\rm air}^{\rm el}}
\def\spaqe{\sigma_{p{-}\rm air}^{q{-}\rm el}}
\def\sigevennu{\sigma_{\rm even}(\nu)}

\newcommand{\ba}{\begin{eqnarray}}
\newcommand{\re}{{\rm Re\,}}
\newcommand{\im}{{\rm Im\,}}
\newcommand{\ea}{\end{eqnarray}}
\newcommand{\be}{\begin{equation}}
\newcommand{\ee}{\end{equation}}

\newcommand{\eq}[1]{Eq.\,(\ref{#1})}

\newcommand{\pbar}{\bar p}

\newcommand{\alphabold}{\mbox{\small\boldmath $\alpha$}}
\newcommand{\xbold}{\mbox{\boldmath $x$}}

\newcommand{\delchisq}{\Delta \chi^2_i(x_i;\alphabold)}
\newcommand{\delchi}{\Delta \chi^2_i}

\newcommand{\delchimax}{{\delchi}_{\rm max}}

\newcommand{\pbarp}{\mbox{$\bar p p$\ }}

\newcommand{\betaP}{\beta_{\cal P'}}

\def\bea{\begin{eqnarray}} 
\def\eea{\end{eqnarray}} 
\begin{document}

\preprint{ANL-HEP-PR-06-79}

\title{Hadronic Cross sections: from cyclotrons to colliders to cosmic rays}

\author{M.~M.~Block}
\affiliation{Department of Physics and Astronomy, Northwestern University, 
Evanston, IL 60208,\\ 
\\{\rm Paper presented at the 2010 ISVHECRI conference, FermiLab, June 28---July 2, 2020 \\
} 
}
\date{\today}

\begin{abstract}
We present evidence for the saturation of the Froissart bound at high energy for {\em all hadronic} total cross sections at high energies, and use this to unify $pp$ (and $\bar p p$) total cross sections over the energy range from cyclotrons to colliders to ultra-high energy cosmic rays, an energy span from $\sqrt s = 4$ GeV to 80 TeV.

\end{abstract}

\maketitle

{\em Introduction.} High energy cross sections for the scattering of hadrons should be bounded   by $\sigma \sim \ln^2s$,
where $s$ is the square of the cms energy.   This fundamental result is derived from unitarity and analyticity by Froissart \cite{froissart}, who states:

``At forward or backward angles, the modulus of the amplitude behaves at most like $s\ln^2s$, as $s$ goes to infinity.  We can use the optical theorem to derive that the total cross sections behave at most like $\ln^2s$, as $s$ goes to infinity".

In this context, saturating the Froissart bound refers to an energy dependence of the total  cross section rising no more rapidly than  $\ln^2s$. 

It will be shown that the Froissart bound is {\em saturated} at high energies  in 
$\gamma p$, $\pi^{\pm}p$ and $\pbarp$ and $pp$ scattering \cite{bh},  as well as in  $\gamma^*p$ scattering \cite{bbt}, as seen from Deep Inelastic Scattering (DIS) in $e+p\rightarrow e+X$. 

Using Finite Energy Sum Rules (FESR) derived from analyticity constraints---in order to anchor accurately cross sections at cyclotron energies \cite{mbFESR}---we will make precise predictions about the total $pp$ cross section, the $\rho$-value (the ratio of the real to the imaginary portion of the forward $pp$ scattering amplitude), as well as the shape of the differential elastic scattering cross section, $d\sigma_{rme el}/dt$, at the LHC.  Further, we will make predictions of the total $pp$ cross section at cosmic ray energies, up to 50 TeV,  and will compare them to the latest experiments.

{\em Data selection.} We make the following major assumptions about the experimental data that we fit:
\begin{enumerate}
\item The experimental data can be  fitted by a model which successfully describes the data.
\item The signal data are Gaussianly distributed, with Gaussian errors.
\item The noise data consists only of points ``far away'' from the true signal, {\em i.e.}, ``outliers'' only.
\item The outliers do not completely swamp the signal data. 
\end{enumerate}

We will use the ``Sieve'' algorithm \cite{mbSieve} to remove ``outliers'' in the cross section and $\rho$-values that we will fit, in order to improve the accuracy of our fits. The ``Sieve'' Algorithm does the following:
 
\begin{enumerate}
\item{Make a robust fit of {\em all} of the data (presumed outliers and all)\ by minimizing $\Lambda^2_0$, the Lorentzian squared, defined as
\ba
\Lambda^2_0(\alphabold;\xbold)\equiv\sum_{i=1}^N\ln\left\{1+0.179\delchisq\right\},\nonumber\\
\quad {\rm where \ }\delchisq\equiv \left(\frac{y_i-y(x_i;\alphabold)}{\sigma_i}\right)^2.
\ea 
The $M$-dimensional parameter space of the fit is given by  $\alphabold=\{\alpha_1,\ldots,\alpha_M\}$; $\xbold=\{{x_1,\ldots,x_N}\}$ is the abscissa of the $N$ experimental measurements $\mbox{\boldmath $y$}=\{y_1,\ldots,y_N\}$ that are  being fit;   $y(x_i;\alphabold)$ is the theoretical value at $x_i$ and $\sigma_i$ is the experimental error.
 Minimizing $\Lambda^2_0$ gives  the same total $\chi^2_{\rm min}\equiv\sum_{i=1}^N \delchisq$  as that found in a $\chi^2$ fit,  as well as  rms widths (errors) for the parameters---{\em for Gaussianly distributed data}---that are almost the same as those found in a $\chi^2$ fit. The quantitative measure of ``far away'' from the true signal, {\em i.e.,} point $i$ is an  outlier,  is the magnitude of its $\delchisq= \left(\frac{y_i-y(x_i;\alphabold)}{\sigma_i}\right)^2$. 

If $\chi^2_{\rm min}$ is satisfactory, make a conventional $\chi^2$ fit to get the errors and you are finished.   If $\chi^2_{\rm min}$ is not satisfactory, proceed to step 
 \ref{nextstep}.} 
\item {Using the above robust $\Lambda^2_0$ fit as the initial estimator for the theoretical curve, evaluate $\delchisq$, for each of the $N$ experimental points.}\label{nextstep}
\item A largest cut, $\delchisq_{\rm max}$, must now be selected. For example, we might start the process with $\delchisq_{\rm max}=9$. If any of the points have $\Delta \chi^2_i(x_i;\alphabold)>\delchisq_{\rm max}$, reject them---they fell through the ``Sieve''. The choice of $\delchisq_{\rm max}$ is an attempt to pick  the largest ``Sieve'' size (largest $\delchisq_{\rm max}$) that rejects all of the outliers, while minimizing the number of signal points  rejected. \label{redo}
\item Next, make a conventional $\chi^2$ fit to the sifted set---these data points are the ones that have been retained in the ``Sieve''. This  fit is used to estimate   $\chi^2_{\rm min}$.    Since the data set has been truncated by eliminating the points with $\delchisq>\delchisq_{\rm max}$, we must slightly renormalize the $\chi^2_{\rm min}$ found to  account for this, by the factor $\cal R$=1.027, 1.14, 1.291 for $\delchimax=9,6,4$. If the renormalized $\chi^2_{\rm min}$, {\em i.e.,} ${\cal R}\times \chi^2_{\rm min}$ is acceptable---in the {\em conventional} sense, using the $\chi^2$ distribution probability function---we consider the fit of the data to the  model to be satisfactory  and proceed to the next step. If the renormalized $\chi^2_{\rm min}$ is not acceptable and $\delchisq_{\rm max}$ is not too small, we pick a smaller 
$\delchisq_{\rm max}$ and go back to step \ref{redo}. The smallest value of $\delchisq_{\rm max}$ that makes much sense, in our opinion, is $\delchisq_{\rm max}>2$.  One of our primary assumptions is that the noise doesn't swamp the signal. If it does, then we must discard the model---we can do nothing further with this model and data set!

\item
{From the  $\chi^2$ fit that was made to the ``sifted'' data in the preceding step, evaluate  the parameters $\alphabold$.
Next, evaluate the $M\times M$ covariance (squared error) matrix of the parameter space found in the $\chi^2$ fit. We find the {\em renormalized}   squared error matrix of our $\chi^2$  fit by multiplying the covariance matrix by the square of the  factor $r_{\chi^2}$ (we find $r_{\chi^2}\sim 1.02,1.05$, 1.11 and 1.14 for $\delchisq_{\rm max}=9$, 6, 4 and 2). The values of $r_{\chi^2}>1$ reflect the fact that a $\chi^2$ fit to the {\em truncated} Gaussian distribution that we obtain has a rms (root mean square) width which is somewhat greater than the  rms width of the $\chi^2$ fit to the same untruncated distribution. Extensive computer simulations  demonstrate that this {\em robust} method of error estimation yields accurate error estimates and error correlations, even in the presence of large backgrounds.}
\end{enumerate}

You are now finished.  The initial robust $\Lambda^2_0$ fit has been used  to allow the phenomenologist to find a    sifted data set. The subsequent application of a $\chi^2$ fit to the {\em sifted set} gives stable estimates of the model parameters $\alphabold$, as well as a goodness-of-fit of the data to the model when $\chi^2_{\rm min}$ is renormalized for the effect of truncation due to the cut $\delchisq_{\rm max}.$   Model parameter errors are found when the covariance (squared error) matrix of the $\chi^2$ fit is multiplied by the appropriate factor $(r_{\chi^2})^2$ for the cut $\delchisq_{\rm max}$. 

{\em Analyticity constraints on hadronic cross sections.} Block and Cahn \cite{bc} used 
an even real analytic amplitude $\tilde f_+(\nu)$  given by 
\ba
\im \tilde f_+(\nu)&=&\frac{p}{4\pi}\left[c_0+c_1\ln\left(\frac{\nu}{m}\right)+c_2\ln^2\left(\frac{\nu}{m}\right)\right.\nonumber\\
&&+\left. \betaP\left(\frac{\nu}{m}\right)^{\mu-1}\right]
\quad{\rm for }\ \nu\ge m,\nonumber\\
\im \tilde f_+(\nu)&=&0\qquad{\rm for }\ 0\le\nu\le m,\label{imagf+bh}\\
\re \tilde f_+(\nu)&=&\frac{p}{4\pi}\left[\frac{\pi}{2}c_1+c_2\pi \ln\left(\frac{\nu}{m}\right)\right.\nonumber\\
&&-\left.\beta_{\cal P'}\cot\left({\pi\mu\over 2}\right)\left(\frac{\nu}{m}\right)^{\mu -1}\right],\label{realf+bh}
\ea
where $\nu$ is the nucleon (pion) laboratory energy and $m$ is the proton mass. 
Using the optical theorem, the even cross section is 
\be
\tilde\sigma_+(\nu)=c_0+c_1\ln(\nu/m)+c_2\ln^2(\nu/m)+\betaP(\nu/m)^{\mu-1},\label{sig+bh}
\ee
where here the coefficients $c_0$, $c_1,\ c_2$ and $\betaP$ have dimensions of mb.  Since $\frac{\nu}{m}=\frac{s}{2m^2}$ for $\nu\gg m$, i.e., for the high energy regime in which we are working, we will henceforth refer to equations similar to that of \eq{sig+bh} as  $\ln^2 s$ fits if $c_2\ne 0$ and as  $\ln s$ fits if $c_2=0$.

We now introduce $f_+(\nu)$, the {\em true} even forward scattering amplitude (which of course, we do not know!), valid for all $\nu$, where $f_+(\nu)\equiv[f_{pp}(\nu)+f_{\bar pp}(\nu)]/2$, using forward scattering amplitudes for $pp$ and $\bar pp$ collisions. 
Using the optical theorem, the imaginary portion of $f_+(\nu)$  is related to the  even  total cross section $\sigevennu$ by
\ba
\im f_+(\nu)&=&\frac{p}{4\pi}\sigevennu \!\qquad\mbox{for $\nu\ge m$}.
\label{sig+1}
\ea

Next, define the  odd amplitude $\nu\hat f_+(\nu)$ as the difference 
\be
\nu\hat f_+(\nu)\equiv\nu\left[f_+(\nu)-\tilde f_+(\nu)\right]\label{fsuper}
,
\ee
which satisfies the unsubtracted odd amplitude  dispersion relation
\be
\re \nu \hat f_+(\nu)=\frac{2\nu}{\pi}\int^\infty_0\frac{\im \nu\,' \hat f_+(\nu\,')}{\nu\,'^2-\nu^2}\,d\nu\,'. \label{dispersion}
\ee
Since for large $\nu$, the odd amplitude $\nu \hat f_+(\nu)\sim\nu^\alpha$ ($\alpha<0$) by design, it also satisfies the super-convergence relation
\be
\int_0^\infty{\rm Im}\,\nu\hat f_+(\nu)\,d\nu=0.\label{superconvergence}
\ee

In Ref. \cite{dolen-horn-schmid}, the FESRs are given by
\be
\int_0^{\nu_0}\nu^n\,{\rm Im}\hat f \,d\nu=\sum \frac {{\nu_0}^{\alpha +n+1}}{\alpha + n+1},\quad n= 0,1,\ldots,\infty, \label{FESRHorn}
\ee
where $\hat f(\nu)$ is crossing-even for odd integer $n$ and crossing-odd for even integer $n$. 
In analogy to the $n=1$ FESR of Ref..  \cite{dolen-horn-schmid}, which requires the odd amplitude $\nu \hat f(\nu)$, Igi and Ishida inserted the super-convergent  amplitude of \eq{fsuper} into the super-convergent dispersion relation of \eq{superconvergence}, obtaining 
\be
\int_0^{\infty}\nu \,\im \left[f_+(\nu)- \tilde f_+(\nu)\right]\, d\nu.\label{super0}
\ee
We note that the odd difference amplitude $\nu \im  \hat f_+(\nu)$ satisfies \eq{superconvergence}, a super-convergent dispersion relation, even if neither $\nu\, \im f_+(\nu)$ nor $\nu \,\im \tilde f_+(\nu)$ satisfies it.  Since the integrand  of \eq{super0}, $\nu \,\im \left[f_+(\nu)- \tilde f_+(\nu)\right]$,  is super-convergent, we can truncate the upper limit of the integration at the finite energy $\nu_0$, an energy high enough for resonance behavior to vanish and where the difference between the two amplitudes---the true amplitude $f_+(\nu)$ minus $\tilde f_+(\nu)$, the amplitude which parametrizes the high energy behavior---becomes negligible, so that the integrand can be neglected for energies greater than $\nu_0$. Thus, after some rearrangement, we get the even finite energy sum rule (FESR)
\be
\int_0^{\nu_0}\nu \im f_+(\nu)\,d\nu=\int_0^{\nu_0}\nu\im \tilde f_+(\nu)\, d\nu.
\label{truef}
\ee 

Next, the left-hand integral of \eq{truef} is broken up into two parts, an integral from $0$ to $m$ (the `unphysical' region) and the integral from $m$ to ${\nu_0}$, the physical  region. We use the optical theorem to evaluate the left-hand integrand for $\nu\ge m$. After noting that the imaginary portion of $\tilde f_+(\nu)=0$ for $0\le\nu\le m$, we again use the optical theorem  to evaluate the right-hand integrand, finally obtaining  the finite energy sum rule FESR(2) of Igi and Ishida  \cite{igi}, in the form: 
\ba
\int_0^m\nu\, \im f_+(\nu)\,d\nu+\frac{1}{4\pi}\int_m^{{\nu_0}}\nu p\,\sigma_{\rm even}(\nu)\,d\nu = \nonumber\\
 \ \ \ \ \ \ \ \frac{1}{4\pi}\int_m^{\nu_0}\nu p\,\tilde\sigma_+(\nu)\, d\nu.\hphantom{\frac{1}{4\pi}\int_m^{{\nu_0}}\nu p\,\sigma_{\rm even}(\nu)\,d\nu}\label{intftilde}
\ea

We now enlarge on the consequences of \eq{intftilde}. We note that if \eq{intftilde} is valid at the upper limit ${\nu_0}$, it certainly is also valid at ${\nu_0}+\Delta {\nu_0}$, where  $\Delta {\nu_0}$ is very small compared to ${\nu_0}$, i.e., $0\le \Delta \nu_0\ll\nu_0$. Evaluating \eq{intftilde} at the energy ${\nu_0}+\Delta {\nu_0}$ and then subtracting \eq{intftilde} evaluated at ${\nu_0}$, we find
\be
\frac{1}{4\pi}\int_{\nu_0}^{{\nu_0}+\Delta {\nu_0}}\!\!\!\!\!\!\nu p\,\sigevennu\,d\nu=\frac{1}{4\pi}\int_{\nu_0}^{{\nu_0}+\Delta {\nu_0} }\!\!\!\!\!\!\nu p\tilde\sigma_+(\nu)\,d\nu.\label{FESR(2)_4}
\ee

Clearly, in the limit of $\Delta {\nu_0}\rightarrow 0$, \eq{FESR(2)_4} goes into
\be
\sigma_{\rm even}({\nu_0})=\tilde\sigma_+({\nu_0}).\label{sig+=sighigh}
\ee
Obviously, \eq{sig+=sighigh} also implies that
\be
\sigma_{\rm even}({\nu})=\tilde\sigma_+({\nu})\qquad\mbox{\rm for all $\nu\ge \nu_0$},\label{sig+=sighigh1}
\ee
but is most useful in practice when $\nu_0$ is as low as possible. The utility of \eq{sig+=sighigh1} becomes evident when we recognize that the left-hand side of it can be evaluated using  the very accurate low energy {\em experimental} crossing-even total cross section data, whereas the right-hand side can use the phenomenologist's parameterization of the {\em high} energy cross section. For example, we could use the cross section parameterization of \eq{sig+bh} on the right-hand side of \eq{sig+=sighigh1} and write the constraint
\ba
\left[\sigma_{pp}(\nu)+\sigma_{\bar pp}(\nu)\right]/2=c_0+c_1\ln(\nu/m)+c_2\ln^2(\nu/m)\nonumber\\
+\betaP(\nu/m)^{\mu-1},\hphantom{xxxxxxxxxx}\label{sig+bh1}
\ea
where $\sigma_{pp}$ and $\sigma_{\bar pp}(\nu)$ are the {\em experimental} $pp$ and $\bar pp$ cross sections at the laboratory energy $\nu$.
Equation (\ref{sig+=sighigh}) (or \eq{sig+=sighigh1}) is our first important extension, giving us  an analyticity constraint, a consistency condition  that the even high energy (asymptotic) amplitude must satisfy.  

Reiterating, \eq{sig+=sighigh1} is a consistency condition imposed by analyticity that states that we must fix the even high energy cross section  evaluated at energy $\nu\ge {\nu_0}$ (using the asymptotic even amplitude) to the low energy {\rm experimental} even cross section at the {\em same}  energy $\nu$, where $\nu_0$ is an energy just above the resonances. Clearly, \eq{sig+=sighigh} also implies that all derivatives of the total cross sections match, as well as the cross sections themselves, i.e., 
\ba
\frac{d^n\sigma_{\rm even}}{d\nu^n\ \ \ \  }({\nu})&=& \frac{d^n\tilde\sigma_+}{d\nu^n\ }({\nu}),\ n=0,1,2,\ldots\ \nu\ge \nu_0,\label{derivative}
\ea
giving new even amplitude  analyticity constraints. Of course, the evaluation of \eq{derivative} for $n=0$ and $n=1$ is effectively the same as evaluating \eq{derivative} for $n=0$ at two nearby values, $\nu_0$ and $\nu_1>\nu_0$. It is up to the phenomenologist to decide which {\em experimental set}\, of quantities it is easier to evaluate.

We emphasize that these consistency constraints  are the consequences of imposing analyticity, implying several important conditions:
\begin{enumerate}
\item 
The new constraints that are derived here tie together both the even  hh and $\bar {\rm h}$h experimental cross sections and their derivatives to the even high energy approximation that is used to fit data at energies well above the resonance region. Analyticity then requires that there  should be a {\em good} fit to the high energy data {\em after} using these constraints, i.e., the $\chi^2$ per degree of freedom of the constrained fit should be $\sim 1$, {\em if}\, the high energy asymptotic amplitude is a  good approximation to the high energy data. This is our consistency condition demanded by analyticity.  If, on the other hand,  the high energy asymptotic amplitude would have given a somewhat poorer fit to the data when {\em not} using the new constraints, the effect is tremendously magnified by utilizing  these new constraints, yielding a very large $\chi^2$ per degree of freedom.  As an example, both Block and Halzen \cite{bh} and Igi and Ishida \cite{igi} {\em conclusively} rule out  a $\ln s$ fit to both $\pi^\pm p$ and $pp$ and $\bar pp$ cross sections and $\rho$-values because it has a huge $\chi^2$ per degree of freedom.\label{point1}
\item  Consistency with analyticity  requires that the results be valid for all $\nu\ge {\nu_0}$, so that the constraint doesn't depend on the particular choice of $\nu$.\label{point2} 
\item The non-physical integral $\int_0^m\nu\, \im f_+(\nu)\,d\nu$ is {\em not} needed for our new constraints. Thus, the value of non-physical integrals, even if very large, does not affect our new constraints.\label{point3}
\end{enumerate} 

Having restricted ourselves so far to 
even  amplitudes, let us now consider odd amplitudes. It is straightforward to show for  odd amplitudes that FESR(odd) implies that
\ba
\frac{d^n\sigma_{\rm odd}}{d\nu^n \ \  }({\nu})\!= \!\frac{d^n\tilde\sigma_-}{d\nu^n }({\nu}),\ n=0,1,2,\ldots,\  \nu \ge \nu_0,\label{oddconstraints}
\ea
where $\tilde\sigma_-(\nu)$ is the odd (under crossing) high energy cross section approximation and $\sigma_{\rm odd}(\nu)$ is the experimental odd cross section.

Thus, we have now derived new analyticity constraints for {\em both
even and odd cross sections}, allowing us to constrain both $\rm{hh}$ and
$\bar{\rm h}$h scattering. Block and Halzen  \cite{bh} expanded upon these ideas, using linear combinations of cross sections and derivatives to anchor {\em both} even  and odd cross sections.  A total of 4 constraints, 2 even and 2 odd constraints, were used by them in their successful $\ln^2s$ fit to $pp$ and $\bar pp$\  cross sections and $\rho$-values, where they first did a local fit to $pp$ and $\bar pp$\ cross sections and their slopes in the neighborhood of $\nu_0=7.59$ GeV (corresponding to $\sqrt s_0=4$ GeV), to determine the experimental cross sections and their first derivatives at which  they anchored their fit. The data they used in the high energy fit were  $pp$ and $\bar pp$ cross sections and $\rho$-values with energies $\sqrt s\ge 6$ GeV. Introducing the even  cross section $\sigma_0(\nu)$, they parameterized the high energy cross sections and $\rho$ values  \cite{bh} as 
\begin{eqnarray}
\sigma_0(\nu)&{\!\!\! =\!\!\! }&c_0+c_1\ln\left(\frac{\nu}{m}\right)+c_2\ln^2\left(\frac{\nu}{m}\right)+\beta_{\cal P'}\left(\frac{\nu}{m}\right)^{\mu -1}\label{sigma0},\\
\sigma^\pm(\nu)&{\!\!\! =\!\!\! }&\sigma_0\left(\frac{\nu}{m}\right)
\pm\  \delta\left({\nu\over m}\right)^{\alpha -1},\label{sigmapm}\\
\rho^\pm(\nu)&{\!\!\! =\!\!\! }&{1\over\sigma^\pm}\left\{\frac{\pi}{2}c_1+c_2\pi \ln\left(\frac{\nu}{m}\right)-\beta_{\cal P'}\cot\left({\pi\mu\over 2}\right)\left(\frac{\nu}{m}\right)^{\mu -1}\right.\nonumber\\
&&\qquad\quad\left.+\frac{4\pi}{\nu}f_+(0)
\pm\  \delta\tan\left({\pi\alpha\over 2}\right)\left({\nu\over m}\right)^{\alpha -1} \right\}\!\!.\label{rhopm}
\end{eqnarray}
We note that the even coefficients $c_0, c_1, c_2$ and $\beta_{\cal P'}$ are the same as those used in \eq{sig+bh1}. The real constant $f_+(0)$ is the subtraction constant  \cite{bc,gilman} required at $\nu=0$ for a singly-subtracted dispersion relation.  They also  used $\mu=0.5$. 
The odd cross section in \eq{sigmapm} is given by
$\delta\left({\nu\over m}\right)^{\alpha -1},\label{odd1}
$
described by two parameters, the coefficient $\delta$ and the Regge power $\alpha<1$, so that the difference cross section between $pp$ and $\bar pp$ vanishes at high energies. 

We now have new analyticity constraints for both even and odd amplitudes. The fits are  anchored  by the experimental cross section data near the transition energy $\nu_0$. These consistency constraints are due to the application of analyticity to finite energy integrals---the analog of analyticity giving rise to traditional dispersion relations when it is applied to integrals with infinite upper limits.

\begin{figure}[h,t,b] 
\begin{center}
\mbox{\epsfig{file=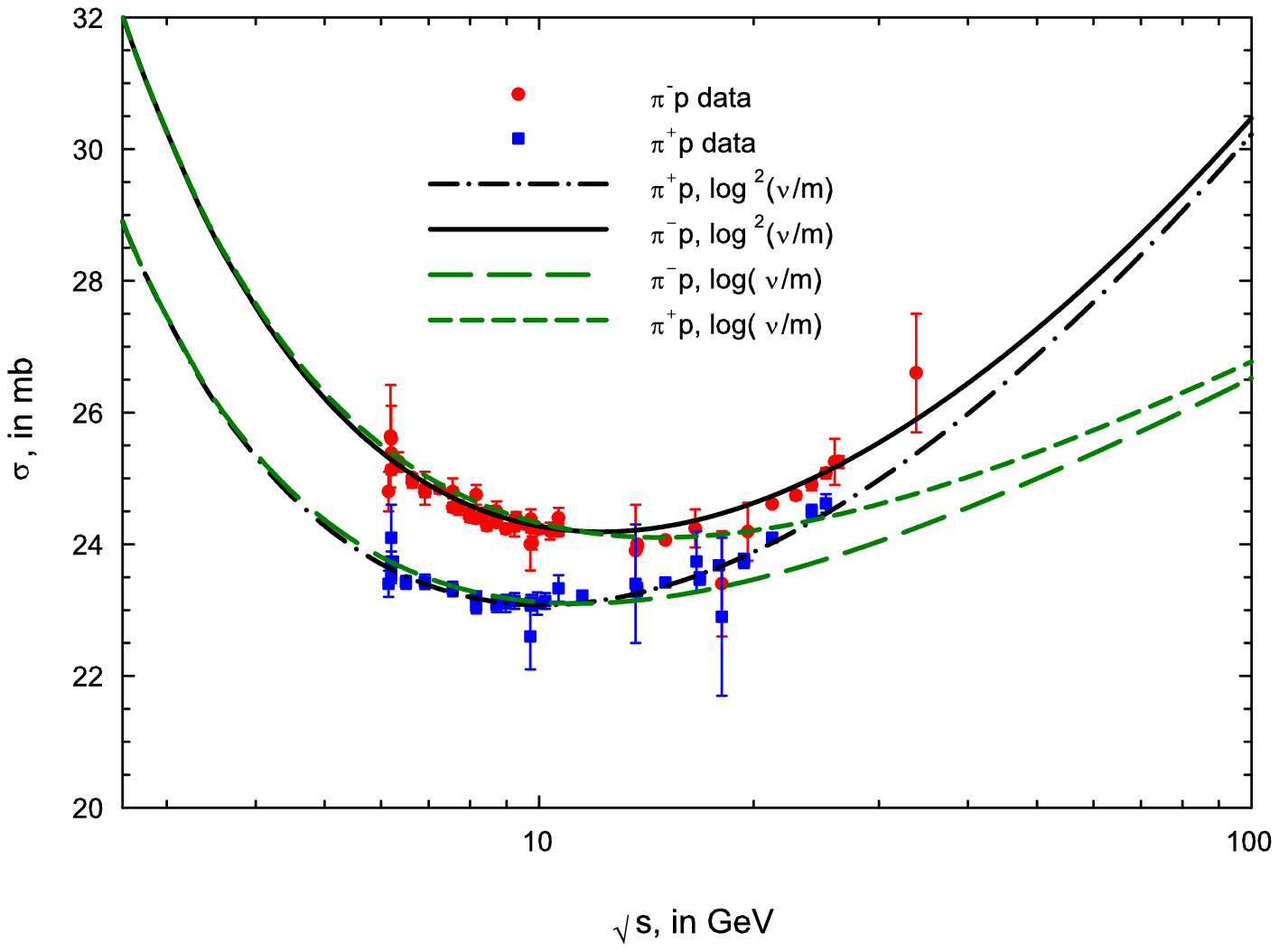
,width=3in,
bbllx=0pt,bblly=0pt,bburx=420pt,bbury=320pt,clip=%
}}
\end{center}
\caption[]{ \footnotesize
The fitted total cross sections $\sigma_{\pi^+p}$ and $\sigma_{\pi^-p}$ in mb, {\em vs.} $\sqrt s$, in GeV, using 4 constraints.  The circles are the sieved  data  for $\pi^-p$ scattering and the squares are the sieved data for $\pi^+p$ scattering for $\sqrt s\ge 6$ GeV. The dash-dotted curve ($\pi^+ p$)  and the solid curve ($\pi^- p$) are $\ln^2s$ fits.  The short dashed curve ($\pi^+ p$) and the long dashed curve ($\pi^- p$) are $\ln s $ fits and clearly do {\em not } fit the data.
}
\label{fig:sigpip}
\end{figure}

\begin{figure}[h,t,b] 
\begin{center}
\mbox{\epsfig{file=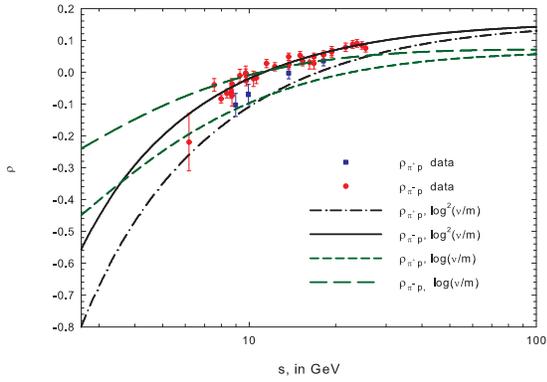,width=3in%
,bbllx=70pt,bblly=290pt,bburx=525pt,bbury=590pt,clip=%
}}
\end{center}
\caption[]{ \footnotesize
The fitted $\rho$-values, $\rho_{\pi^+p}$ and $\rho_{\pi^-p}$, {\em vs.} $\sqrt s$, in GeV, using  4 constraints.  The circles are the sieved data  for $\pi^-p$ scattering and the squares are the sieved data for $\pi^+p$ scattering for $\sqrt s\ge 6$ GeV. The  dash-dotted curve ($\pi^+ p$) and the solid curve ($\pi^- p$) are $\ln^2 s$ fits.  The short dashed curve ($\pi^+ p$) and the long dashed curve ($\pi^- p$)  are $\ln s$ fits, which again are bad fits to the experimental data.
\label{fig:rhopip}
}
\end{figure}

{\em $\gamma p$, $\pi^\pm p$, $\bar p p$ and $pp$ scattering.} Using the 4 constraints of the previous Section, we show in Fig. \ref{fig:sigpip} both the $\ln^2s$ fit and the $\ln s$ fit of \eq{sigma0} and \eq{sigmapm} to the experimental $\pi^\pm p$ total cross sections. The high energy fit is anchored to the very accurate data a at $\sqrt s= 2.6$ GeV. Because of the 4 constraints, the  $\ln^2s$ fit needs only to fit the two parameters $c_1$ and $c_2$, whereas the $\ln s$ fit sets $c_2=0$ and fits only $c_1$. We see that the $\ln^2s$ fit, which {\em saturates the Froissart bound}, gives an excellent fit to the data, while the $\ln s$ fit is ruled out. Shown in Fig. \ref{fig:rhopip} are the  $\rho$-values for $\pi^\pm p$ scattering: again, we see that the $\ln s$ fit is ruled out, whereas we get a good fit when we saturate the Froissart bound.

\begin{figure}[h,t,b] 
\begin{center}
\mbox{\epsfig{file=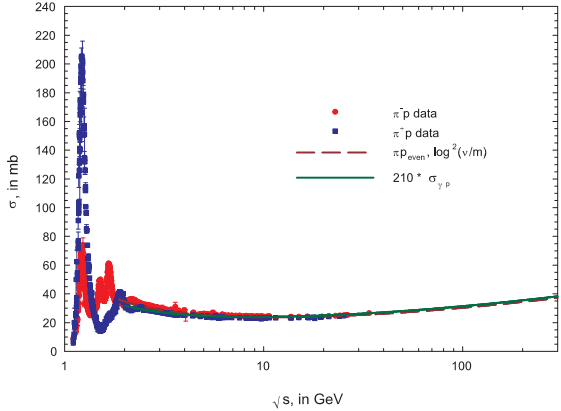,width=3in%
,bbllx=0pt,bblly=0pt,bburx=420pt,bbury=305pt,clip=%
}}
\end{center}
\caption[]{ \footnotesize
The circles are the cross section data  for $\pi^-p$ scattering and the squares are the cross section data for $\pi^+p$ scattering, in mb, {\em vs.} $\sqrt s$, in GeV, for all of the known data.  The dashed curve is the $\ln^2 s$ $\pi p$ fit to the high energy cross section data  of the even amplitude cross section, $\sigma_0$.  The solid curve is the fit of the  $\gamma p$ cross section data , for  cms energies $\sqrt s\ge2.01$ GeV, whereas the $\pi p$ data (cross sections and $\rho$-values) were fit for cms energies $\sqrt s\ge6$ GeV. The two fitted curves are virtually indistinguishable in the energy region $2\le \sqrt s\le 300$ GeV.
}
\label{fig:sigpipallenergies}
\end{figure}

In Fig. \ref{fig:sigpipallenergies} we compare the results of our fitted $\pi p$ cross section $\sigma_0$ (from \eq{sigma0}) with a rescaled version of $\sigma_0(\gamma p)$ that was obtained from a fit to all known high energy $\gamma p$ cross sections.  From about $2\le \sqrt s\le 300$ GeV, the two saturated $\ln^2 s$ fits are virtually indistinguishable.  Thus we conclude that high energy $\gamma p$ total cross sections also go as $\ln^2 s$.


\begin{figure}[h,t,b] 
\begin{center}
\mbox{\epsfig{file=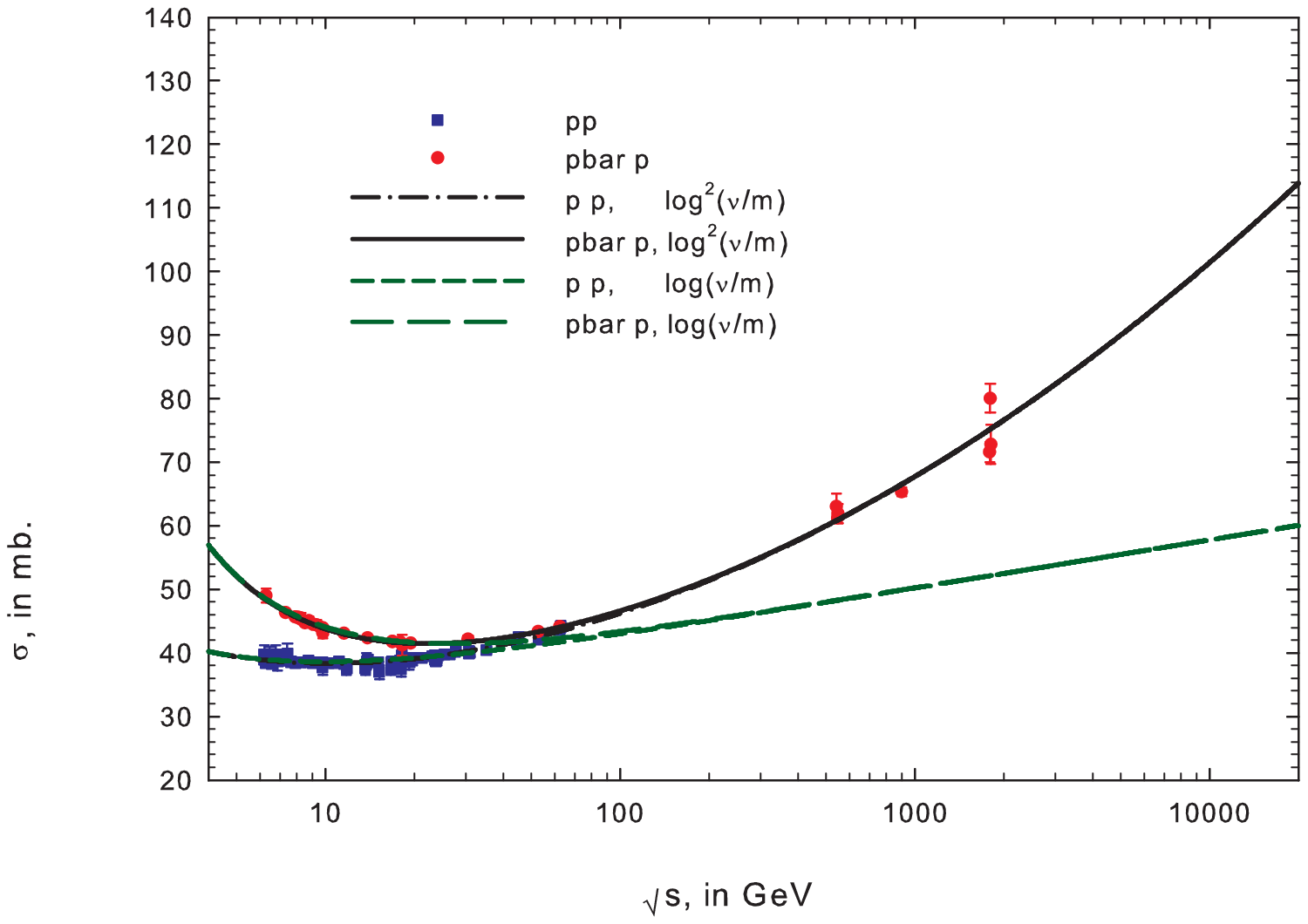,width=3in%
,bbllx=0pt,bblly=0pt,bburx=445pt,bbury=304pt,clip=%
}}
\end{center}
\caption[]{ \footnotesize
The fitted total cross sections $\sigma_{p p}$ and $\sigma_{\pbar p}$ in mb, {\em vs.} $\sqrt s$, in GeV, using  4 constraints. The circles are the sieved data  for $\pbar p$ scattering and the squares are the sieved data for $p p$ scattering for $\sqrt s\ge 6$ GeV. The dash-dotted curve ($pp$)  and the solid curve ($\pbar p$) are $\ln^ s$ fits to  the high energy data.  The short dashed curve ($p p$) and the long dashed curve ($\pbar p$) are $\ln s$ fits to   the high energy data. Clearly, the $\ln s$ fits do {\em not} fit the experimental data.
}
\label{fig:sigmapp}
\end{figure}

\begin{figure}[h,t,b] 
\begin{center}
\mbox{\epsfig{file=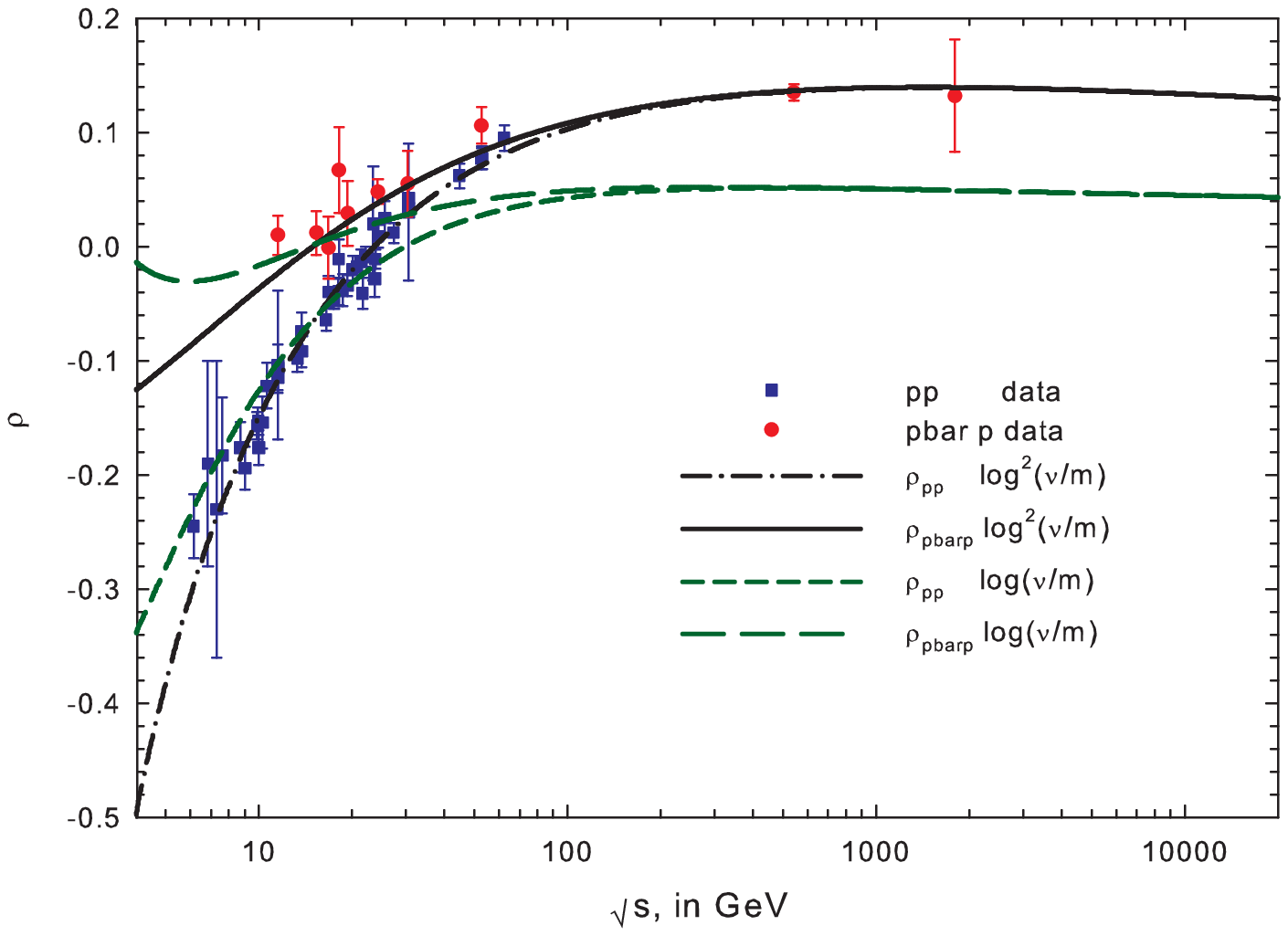,width=3in%
,bbllx=0pt,bblly=0pt,bburx=435pt,bbury=305pt,clip=%
}}
\end{center}
\caption[]{ \footnotesize
The fitted $\rho$-values, $\rho_{p p}$ and $\rho_{\pbar p}$, {\em vs.} $\sqrt s$, in GeV, using  4 constraints. The circles are the sieved data  for $\pbar p$ scattering and the squares are the sieved data for $p p$ scattering for $\sqrt s\ge 6$ GeV. The  dash-dotted curve ($p p$) and the solid curve ($\pbar p$) are $\ln^2 s$ fits  The short dashed curve ($p p$) and the long dashed curve ($\pbar p$)  are $\ln s$ fits. Clearly, the $\ln s$ fits do {\em not} fit the experimental data.
  }
\label{fig:rhopp}
\end{figure}

We next consider high energy $\bar pp$ and $pp$ scattering. Shown in Fig. \ref{fig:sigmapp} are fits to the total cross sections, where again, for the $\ln^2 s$ fit,  {\em only 2} parameters, $c_1$ and $c_2$ are fit, after anchoring the high energy data to the accurate cyclotron 
 data at $\sqrt s= 4$ GeV. Once again, we see that we have an excellent $\ln^2s$ fit to all of the high energy total  cross section data when we anchor our fit (of only 2 parameters!) to the low energy data, whereas the $\ln s$ fit to these data fails. In Fig. \ref{fig:rhopp} we compare the nucleon-nucleon $\rho$-values to our 3 parameter fit of $c_0,\ c_1$ and the subtraction constant $f_+(0)$.
%
%
%
%

Before employing the  ``Sieve'' algorithm on the totality of 212 high energy $\bar pp$ and $pp$ cross sections, the $\chi^2/d.f.$ was 5.7 for the $\ln^2s$ fit, clearly an unacceptably high value. After ``sieving'', the renormalized $\chi^2/d.f.$ was 1.09, for 184 degrees of freedom, using a $\Delta \chi^2_i>6$ cut.  The total $\chi^2$ was 201.4, corresponding to a probability of fit $\approx 0.2$. In all, the 25 rejected points contributed 981 to the total $\chi^2$, i.e., an {\em average} $\Delta\chi^2$ of about 39 per point!

For the LHC , the $\ln^2s$ fits shown in Fig. \ref{fig:sigmapp} and Fig. \ref{fig:rhopp} predict \ba \sigma_{pp}&=&107.3\pm1.2\ {\rm mb},\ \sqrt s= 14\ {\rm TeV}\\
\rho_{pp}&=&0.132\pm0.001,\ea
 where the quoted errors are due to the uncertainties in the fit parameters.  

\begin{figure}[h,t] 
\begin{center}
\mbox{\epsfig{file=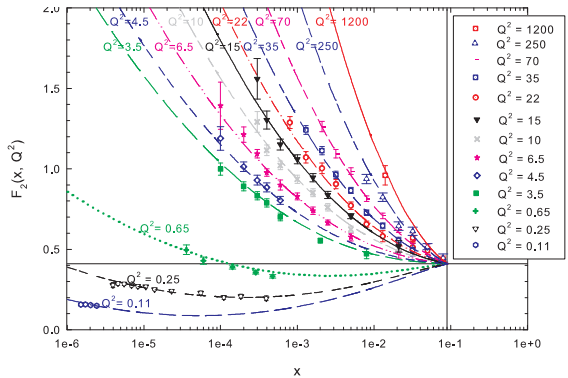,width=3in%
,bbllx=0pt,bblly=390pt,bburx=450pt,bbury=700pt,clip=%
}}
\end{center}
\caption[]{ 
Fits to the proton structure function data, $F_2^p(x, Q^2)$ vs. $x$ of the form $\ln^2s$, for 13 values of $Q^2$.  
The data are from the ZEUS collaboration~  \cite{zeus}. The curves show 13 of 28 global fits   \cite{bbt}.
The vertical and horizontal straight lines intersect at the 
scaling point $x_{\rm P}=0.09,F_2^p(x_{\rm P})=0.41$.  \label{fig:Fvsx}}
\end{figure}

{\em Deep inelastic scattering (DIS).} Berger, Block and Tan  \cite{bbt} analyzed the $x$ dependence of the DIS proton 
structure functions $F^p_2(x, Q^2)$ by beginning with the assumption that the 
$x$ dependence at extremely small x should manifest a behavior consistent 
with saturation of the Froissart bound on hadronic total cross 
sections~ \cite{froissart}, as is satisfied by data on 
$\gamma p$, $\pi^{\pm} p$, and $\bar pp$ and $pp$ interactions. They treated DIS $ep$ scattering as the hadronic reaction $\gamma^*p$, demanding that the total cross section $\sigma(\gamma^*p)$ saturate the Froissart bound of $\ln^2s$. For the process $\gamma^* +p\rightarrow X$, the invariant $s$ is given by $s=Q^2/x$, for small $x$, when $s\gg m^2$, where $m$ is the proton mass, $x$ is the fractional longitudinal momentum of the proton carried by its parton constituents and $Q^2$ is the virtuality of the $\gamma^*$.
Thus, saturting the Froissart  bound \cite{froissart} demands 
that $F_2^p(x,Q^2)$ grow no more rapidly than $\ln^2(1/x)$ at very small $x$. Over the ranges of $x$ and $Q^2$ for which DIS data 
are available, they show that a very good fit to the $x$ dependence of ZEUS 
data  \cite{zeus} is obtained for 
$x \le x_P = 0.09$ and ${Q^2\over x}\gg m^2$ using the expression  
\ba
F_2^p(x,Q^2)=(1-x)&\!\!\!\times&\!\!\!\!\Big\{{F_P\over{1-x_P}}+A(Q^2)\ln\left[\frac {x_P}{x}\frac{1-x}{1-x_P}\right] \nonumber\\
& \!\!\!+ & 
B(Q^2)\ln^2\left[\frac {x_P}{x}\frac{1-x}{1-x_P}\right]\Big\}. \label{Fofx} 
\ea
Their fits to DIS data \cite{zeus} at 24 values of 
$Q^2$ cover the wide range $0.11\le Q^2\le 1200$ GeV$^2$.  The value $x_P = 0.09$ is 
a scaling point  such that the curves for all $Q^2$ pass through the 
point $x = x_P$, at which $F_2(x_P, Q^2) = F_P \sim 0.41$, further constraining all of 
the fits. Figure \ref{fig:Fvsx} shows that also in $\gamma^* p$ scattering, the Froissart bound appears to be saturated.

\begin{figure}[tbp] 
\begin{center}
\mbox{\epsfig{file=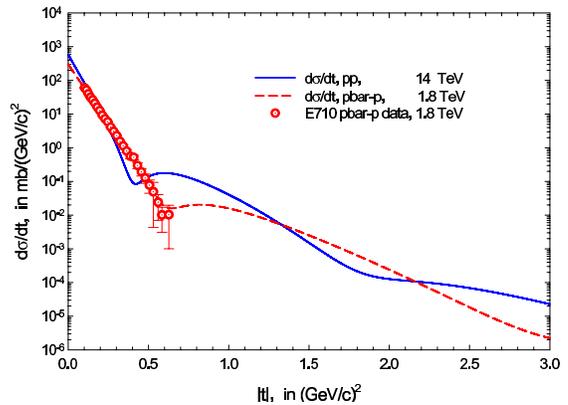,%
		width=3in,bbllx=70pt,bblly=245pt,bburx=500pt,bbury=545pt,clip=%
}}
\end{center}
\caption[$\frac{d\sigma}{dt}$ at the Tevatron and the LHC, using an Aspen Model fit]
{ \footnotesize
The elastic differential scattering cross section 
$\frac{d\sigma}{dt}$, in mb/(GeV/c)$^2$  vs. $|t|$, in (GeV/c)$^2$, using a constrained Aspen Model fit (QCD-inspired theory). 
 The solid curve is the prediction for the reaction $ pp\rightarrow pp$ at the LHC, at $\sqrt s=14$ TeV.
The dashed curve is the prediction for the reaction $\bar pp\rightarrow\bar pp$ at $\sqrt{s}=1.8$ TeV, at the Tevatron Collider; the data 
points are from the E710 experiment. }
\label{fig:dsdt}
\end{figure}

{\em The ``Aspen'' model.}  The ``Aspen'' model is a QCD-inspired eikonal model  \cite{aspen} which naturally yields  total $pp$ and $\bar pp$ cross sections that go as $\ln^2s$ at large energies, i.e., they automatically saturate the Froissart bound. From the model, one can extract the total cross section $\sigma_{\rm tot}$, the $\rho$-value, the elastic cross section $\sigma_{\rm el}$, the nuclear slope parameter $B$ (the logarithmic derivative of $d\sigma_{\rm el}/dt$ at $t=0$) and $d\sigma_{\rm el}/dt$ as a function of $t$, the squared 4-momentum transfer.  Shown in Fig. \ref{fig:dsdt} is a constrained Aspen-model fit  \cite{reports} to  $d\sigma_{\rm el}/dt$ vs. $|t|$, for 1.8 TeV , compared to E710 data at 1.8 TeV; the fit is excellent. Also shown is the prediction for the LHC, at 14 TeV. For more details af the model and for results for constrained total cross sections, etc., see Ref. \cite{reports}.

{\em Cosmic ray predictions.} There are now available published  \cite{fly,akeno,yakutsk,belov} and preliminary  \cite{eastop,ARGO} p-air inelastic production cross sections ($\spai$) that span the enormous $pp$ cms (center-of-mass system) energy range $0.1\la \sqrt s\la 100$ TeV, reaching energies  well above the Large Hadron Collider (LHC). Further,  we expect high statistics results from the Pierre Auger Collaboration  \cite{auger} in the near future in this ultra-high energy region.  Most importantly, we now have available very accurate predictions at cosmic ray energies for the total $pp$ cross section, $\sigma_{pp}$,  from fits   \cite{blockhalzenpp} to accelerator data that used adaptive data sifting algorithms  \cite{sieve} and analyticity constraints{  \cite{blockanalyticity} that were not available in the earlier work of Block, Halzen and Stanev  \cite{blockhalzenstanev}. Here we take advantage of these new $pp$ cross section predictions in order to make {\em accurate} predictions of the cosmic ray p-air total cross sections, to be  compared with past and future experiments.

Extracting proton--proton cross sections from published cosmic ray observations of extensive air showers, and vice versa, is far from
 straightforward  \cite{engel}. By a variety of experimental techniques,
 cosmic ray experiments map the atmospheric depth at which extensive air 
 showers develop and measure the distribution of $X_{\rm max}$, the shower maximum, which is sensitive to the inelastic p-air cross section $\spai$. From the measured $X_{\rm max}$ distribution, the experimenters deduce $\spai$. We will compare published  \cite{fly,akeno,yakutsk} and recently announced  preliminary values of $\spai$ with predictions made from  $\sigma_{pp}$, using a Glauber model to obtain $\spai$ from $\sigma_{pp}$.
  
{\em $\spai$ from the $X_{\rm max}$ distribution: Method I.} The measured shower attenuation length ($\Lambda_m$) is not only
 sensitive to the interaction length of the protons in the atmosphere
 ($\lpa$)  \cite{myfootnote}, with
\begin{equation}
\Lambda_m = k \lpa = k { 14.4 m_p \over \spai}=k\frac{24,100}{\spai} \,,  \label{eq:Lambda_m}
\end{equation}
(with $\Lambda_m$ and $\lpa$ in g\,cm$^{-2}$, the proton mass $m$ in g, and the inelastic production cross section $\spai$ in mb),  but also depends on the rate at which the energy of the primary proton
 is dissipated into electromagnetic shower energy observed in the
 experiment. The latter effect is parameterized in Eq.\,(\ref{eq:Lambda_m})
 by the parameter $k$. The value of $k$ depends critically on the inclusive
 particle production cross section and its energy dependence in nucleon and meson interactions
 on the light nuclear target of the atmosphere (see Ref.   \cite{engel}). We emphasize that the goal of the cosmic ray experiments is $\spai$ (or correspondingly, $\lpa$), whereas 
 in Method I, the {\em measured} quantity is $\Lambda_m$. Thus,  
a significant drawback of Method I is that one needs a model of
 proton-air interactions to complete the loop between the measured
 attenuation length $\Lambda_m$ and the cross section $\spai$,
 {\em i.e.,} one needs the value of $k$ in Eq. (\ref{eq:Lambda_m}) to compute $\spai$. 
Shown in Table \ref{ktable} are the widely varying values of $k$  used in the different experiments.  Clearly the large  range of $k$-values, from 1.15 for EASTOP  \cite{eastop} to 1.6 for Fly's Eye  \cite{fly} differ significantly, thus making the {\em published} values of $\spai$ unreliable.  It is interesting to note the monotonic decrease over time in the $k$'s used in the different experiments, from 1.6 used in Fly's Eye in 1984 to the 1.15 value used in EASTOP in 2007, showing the time evolution of Monte Carlo models of energy dissipation in showers.  For comparison, Monte Carlo simulations  made by Pryke  \cite{pryke} in 2001 of  several more modern shower models are also shown in Table \ref{ktable}. Even among modern shower models, the spread is still significant: one of our goals is to minimize the impact of model dependence on the $\spai$ determination.
 
\begin{table}
\caption{A table of $k$-values, used in experiments and from Monte Carlo model simulation\label{ktable}} 
 \begin{tabular}[b]{|l||c|}
\hline
Experiment&k\\ \hline\hline
Fly's Eye&1.6\\ \hline
AGASA&1.5\\ \hline
Yakutsk&1.4\\ \hline
EASTOP&1.15\\ \hline

     \multicolumn{2}{c}{Monte Carlo Results: C.L. Pryke}\\ 
      \hline
     Model&$k$ \\ \hline
     CORSIKA-SIBYLL&$1.15\pm 0.05$\\
          \hline
MOCCA--SIBYLL&$1.16\pm 0.03$\\
          \hline\hline
CORSIKA-QGSjet&$1.30\pm 0.04$\\
          \hline
MOCCA--Internal&$1.32\pm 0.03$\\
\hline\hline
\end{tabular}
\end{table}

{\em $\spai$ from the $X_{\rm max}$ distribution: Method II.} The HiRes group  \cite{belov} has developed a quasi model-free method of measuring $\spai$ directly.  They fold into their shower development program a randomly generated exponential distribution of shower first interaction points, and then fit the entire distribution, and not just the trailing edge, as  done in  other experiments  \cite{fly,akeno,yakutsk,eastop}.  They obtain $\spai=460\pm14\ ({\rm stat})+39\ ({\rm syst})-11\ ({\rm syst})$ mb at $\sqrt s=77$ TeV, which they claim is effectively model-independent and hence is an absolute determination  \cite{belov}.

{\em Extraction of $\sigma_{pp}$ from $\spai$}. The total $pp$ cross section is extracted from $\spai$ in two distinct steps. First, one calculates the $p$-air {\em total} cross section, $\spa$,  from
 the measured inelastic production cross section using 
\be
\spai = \spa - \spae - \spaqe \,. \label{eq:spa}
\ee
Next, the Glauber method \cite{yodh} is used to transform the measured
 value of $\spai$ into the $pp$  total cross section $\sigma_{pp}$;
 all the necessary steps are calculable in the theory. In Eq.\,(\ref{eq:spa})
 the measured cross section for particle production is supplemented with
$\spae$ and $\spaqe$, 
 the elastic and quasi-elastic cross section, respectively, as calculated by the
 Glauber theory, to obtain the total cross section $\spa$. The subsequent
 relation between $\spai$ and $\sigma_{pp}$ critically involves the nuclear slope parameter $B$,  the logarithmic slope of forward elastic $pp$ scattering, ${d\sigma_{pp}^{\rm el}/ dt}$. 
 A plot of $B$ against
 $\sigma_{pp}$, 5 curves of different values of $\spai$, is shown in Fig.\,\ref{fig:p-air}, taking into account inelastic screening{ \cite{engelBsig}. The reduction procedure
 from $\spai$ to $\sigma_{pp}$ is summarized in Ref.  \cite{engel}.
 The solid curve in Fig.\,\ref{fig:p-air} is a plot of $B$  vs. 
 $\sigma_{pp}$---with $B$ taken from the ``Aspen'' eikonal model and $\sigma_{pp}$ taken from the $\ln^2 s$ fit using analytic amplitudes. The large dot corresponds to the value of $\sigma_{pp}$ and $B$ at $\sqrt s$ = 77 TeV, the HiRes energy, thus fixing the HiRes predicted value of $\spai$.
\begin{figure}
\begin{center}
\mbox{\epsfig{file=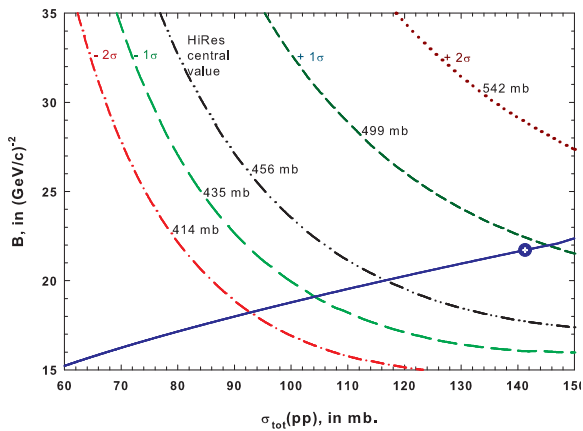%
              ,width=3in,bbllx=2pt,bblly=33pt,bburx=408pt,bbury=340pt,clip=%
}}
\end{center}
\caption[]{  $B$ dependence on the $pp$ total cross section $\sigma_{pp}$. The
 five curves are lines  of constant  $\spai$,  of 414, 435, 456, 499 and
 542 mb---the central value is the published Fly's Eye value, and the others
 are $\pm 1\sigma$ and $\pm 2\sigma$. The solid curve is a plot of a 
 QCD-inspired fit of $B$ against $\sigma_{pp}$, obtained from a $\ln^2s$ fit.  The large dot is the prediction for
$\spai$ at $\sqrt s=77$ TeV, the HiRes energy.}
\label{fig:p-air}
\end{figure}

 {\em Obtaining $\sigma_{pp}$ from  $\spai$}. In Fig.\,\ref{fig:sigpp_p-air}, we have plotted the values of
 $\sigma_{pp}$ vs. $\spai$ that are deduced from the
 intersections of the $B$-$\sigma_{pp}$ curve  with the $\spai$
 curves in Fig.\,\ref{fig:p-air}. Figure~\,\ref{fig:sigpp_p-air}
furnishes cosmic ray experimenters with an easy method to convert their measured $\spai$ to $\sigma_{pp}$, and vice versa. %

\begin{figure}[h]
\begin{center}
\mbox{\epsfig{file=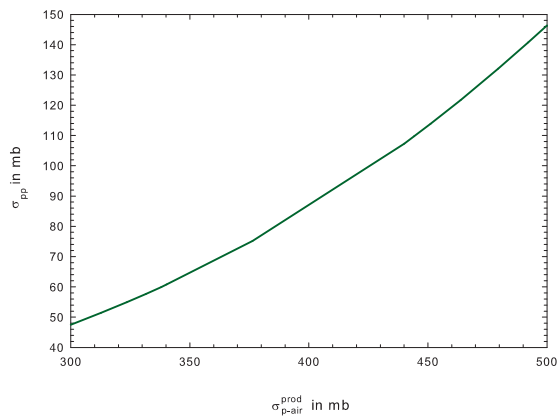%
            ,width=3in,bbllx=65pt,bblly=245pt,bburx=500pt,bbury=560pt,clip=%
}}
\end{center}
\caption[]{A plot of the predicted total pp cross section $\sigma_{pp}$, in mb
 {\em vs.} the predicted production p-air cross section, $\spai$, in mb, where 
}
\label{fig:sigpp_p-air}
\end{figure}

{\em Determining the $k$ value}.  In Method I, the extraction of $\lpa$ (or $\spai$) from the
 measurement of $\Lambda_m$ requires knowing the parameter
 $k$. The measured depth $X_{\rm max}$ at which a shower reaches
 maximum development in the atmosphere, which is the basis of the
 cross section measurement in Ref.  \cite{fly}, is a combined measure
 of the depth of the first interaction, which is determined by
 the inelastic cross section, and of the subsequent shower development,
 which has to be corrected for. 
 The model dependent rate of shower development and its fluctuations
 are the origin of the deviation of $k$ from unity
 in Eq.\,(\ref{eq:Lambda_m}). As seen in Table \ref{ktable}, its values range from 1.6 for a very old model
 where the inclusive cross section exhibited Feynman scaling, to 1.15
 for modern models with large scaling violations.

 Adopting the same  strategy that earlier had been used by Block, Halzen and Stanev \cite{blockhalzenstanev}, we matched the data to our prediction of $\spai (s)$, extracting 
 a {\em common} value for $k$, neglecting the possibility
of a weak energy dependence of $k$ over the range measured, found to be very small in the simulations of Ref.  \cite{pryke}.
By combining the results of $B$ from the ``Aspen'' model and Fig.\,\ref{fig:sigpp_p-air}, we obtain our prediction of $\spai$  vs. $\sqrt s$, which is shown in Fig. \ref{fig:p-aircorrected2}. Leaving $k$ as a free parameter,we make a $\chi^2$ fit to {\em  rescaled} $\spai (s)$ values of Fly's Eye,  \cite{fly}AGASA \cite{akeno}, EASTOP \cite{eastop} and Yakutsk \cite{yakutsk},  the experiments that need a common $k$-value. 

Figure \,\ref{fig:p-aircorrected2} is a plot of $\spai$ vs. $\sqrt s$, the cms energy in GeV, for the two different types of experimental extraction, using Methods I and II described earlier.  Plotted {\em as published} is the HiRes value at $\sqrt s=77$ TeV, since it is an absolute measurement. We have rescaled in Fig.\,\ref{fig:p-aircorrected2} the published values of $\spai$ for  Fly's Eye  \cite{fly}, AGASA  \cite{akeno}, Yakutsk  \cite{yakutsk} and EAS-TOP  \cite{eastop},  against our prediction of $\spai$, using the {\em common} value of $k=1.264 \pm 0.033\pm 0.013$ obtained from a $\chi^2$ fit, and it is the rescaled values that are plotted in Fig. \ref{fig:p-aircorrected2}, along with the rescaled values of ARGO-YBJ  \cite{ARGO}, which were not used in the fit.
The error in $k$ of $0.033$ is the statistical error of the $\chi^2$ fit, whereas the error of $0.013$ is the systematic error due to the error in the prediction of $\spai$. %
\begin{figure}[b]
\begin{center}
\mbox{\epsfig{file=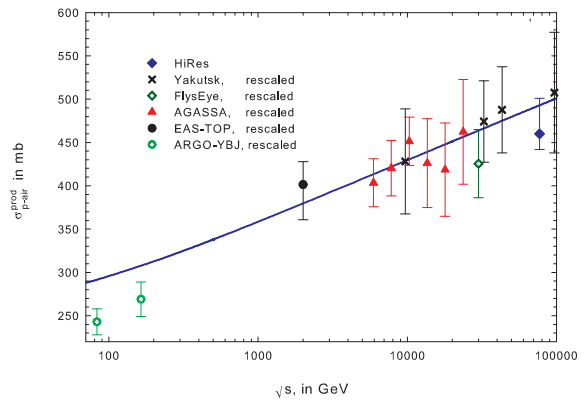%
              ,width=3in,bbllx=0pt,bblly=55pt,bburx=440pt,bbury=360pt,clip=%
}}
\end{center}
\caption[]{\protect
{  A $\chi^2$ fit of the {\em renormalized} AGASA, EASTOP, Fly's Eye and Yakutsk data for $\spai$, in mb,
 as a function of the energy, $\sqrt s$, in GeV. The result of the fit for the
 parameter $k$ in Eq. (\ref{eq:Lambda_m}) is $k=1.263\pm0.033$. The HiRes point (solid diamond), at $\sqrt s=77$ GeV, is model-independent and has {\em not} been renormalized.  The renormalized ARGO-YBJ data were not used in the fit.
}
}
\label{fig:p-aircorrected2}
\end{figure}
%
Clearly,
 we have an excellent fit, with complete agreement for  all experimental points. Our analysis gave  $\chi^2=3.19$ for 11 degrees of freedom (the low
 $\chi^2$ is likely due to overestimates of experimental errors).
 We note that our $k$-value,  $k=1.264\pm0.033\pm0.013$, although somewhat too small for the very low energy ARGO-YBJ data, is about halfway between the values of CORSIKA-SIBYLL and CORSIKA-QSGSjet found in the Pryke simulations  \cite{pryke}, as seen in Table \ref{ktable} for model predictions. We next compare our measured $k$ parameter with a recent   direct measurement of $k$ by the HiRes group  \cite{belovkfactor}. They measured the exponential slope of the tail of their $X_m$ distribution, $\Lambda_m$ and compared it to the p-air interaction length $\lpa$ that they found. Using \eq{eq:Lambda_m}, they deduced that a preliminary value, $k=1.21+0.14-0.09$, in agreement with our value. Taken together with the goodness-of-fit of our fitted $k$ value and the fact that our value, $k=1.264 \pm 0.033\pm 0.013$ is compatible with the range of $k$ values from theoretical models shown in Table \ref{ktable}, the preliminary HiRes value \cite{belovkfactor} of $k=1.21+0.14-0.09$  is additional experimental confirmation of our overall method for determining $k$, i.e.,  our assumptions that the $k$ value is essentially energy-independent, as well as being independent of  the very different experimental techniques for measuring air-showers. Our measured $k$ value, $k=1.264 \pm 0.033\pm 0.013$, agrees very well with the preliminary $k$-value measured by the HiRes group, at the several parts per mil level; in turn, they both agree with Monte Carlo model simulations at the 5--10 part per mil level.  

It should be noted that the preliminary EASTOP measurement \cite{eastop}---at the  cms energy $\sqrt s=2$ TeV---is at an energy essentially  identical to the top energy of the Tevatron collider, where there is an {\em experimental} determination of $\sigma_{\bar pp}$ \cite{comment}, and consequently, no necessity for an {\em extrapolation} of collider cross sections.  Since  their value of $\spai$  is in excellent agreement with the predicted value of $\spai$, this  anchors our fit at its low energy end. Correspondingly, at the high end of the cosmic ray spectrum, the absolute value of the HiRes experimental value of $\spai$ at 77 TeV---which requires {\em no knowledge} of the $k$ parameter---is also in good agreement with our prediction, anchoring the fit at the high end. Thus, our $\spai$ predictions, which span the enormous energy range, $0.1\la\sqrt s\la 100$ TeV, are consistent with cosmic ray data, for both magnitude and energy dependence. 

Shown in Fig. \ref{fig:hirespp} are {\em all} of the known $pp$ and $\bar pp$ totl cross section data, including the cosmic ray points, spanning the energy region from 2 GeV to 80 TeV, fitted by the {\em same}  $\ln^2 s$ (saturated Froissart bound) fit of the even cross section $\sigma_0=(\sigma_{pp}+\sigma_{\bar pp})/2$.

{\em CCCR: Cyclotrons to Colliders to Cosmic Rays.} We have finally reached our goal.  This long energy tale of accelerator experiments, extending over some 55 years,  from those using cyclotrons to  those using synchrotons  and then, finally, to those using colliders,  has now  been unified with those experiments using high energy cosmic rays as their beams.  The accelerator experiments had large fluxes and accurate energy measurements, allowing for precision measurements: the more precise, the lower the energy. On the other hand, the cosmic ray experiments always suffered from low fluxes of particles and poor  energy determinations of their events, but made up for that by their incredibly high energies.   

The ability to clean up accelerator cross section and $\rho$-value data by the Sieve algorithm, along with new fitting techniques using analyticity constraints  in the form of anchoring high energy cross section fits to the value of low energy $pp$ and $\bar pp$ experimental cross sections (and their energy derivatives) have furnished us with a precision fit---using the a $\ln^2s$  form that saturates the Froissart Bound---which allows us to make accurate extrapolations into the LHC and cosmic ray regions, extrapolations guided by the principles of analyticity and unitarity embodied in the Froissart Bound.

\begin{figure}[tbp]
\begin{center}
\mbox{\epsfig{file=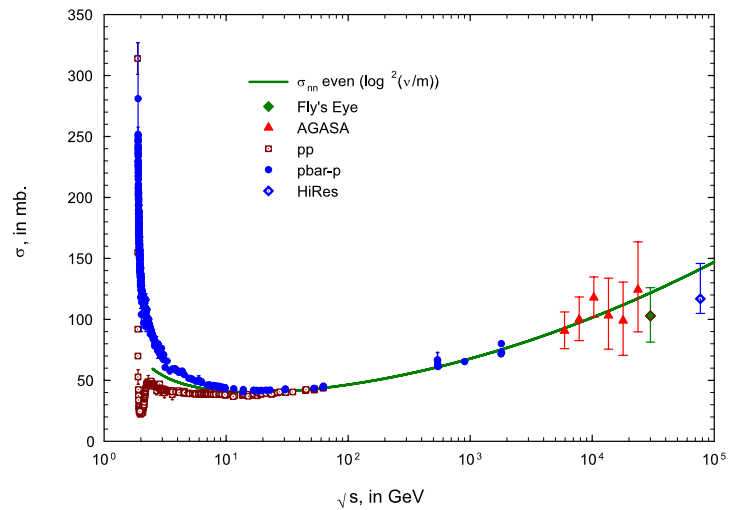%
              ,width=3.8in,bbllx=0pt,bblly=0pt,bburx=430pt,bbury=300pt,clip=%
}}
\end{center}
\caption[All known  \cite{pdg} $\sigma_{pp}$ and $\sigma_{\bar pp}$ accelerator total cross sections, shown  together with $\sigma_{pp}$ deduced from the AGASA, Fly's Eye and HiRes cosmic ray experiments]
{\footnotesize
\protect
{ All known  \cite{pdg} $\sigma_{pp}$ and $\sigma_{\bar pp}$ accelerator total cross sections, shown  together with $\sigma_{pp}$ deduced from the AGASA, Fly's Eye and HiRes cosmic ray experiments. $pp$ and $\bar pp$  accelerator total cross sections, in mb,  vs. the c.m. energy $\sqrt s$, in GeV. The circles are $\bar pp$ and the open squares are $pp$ data. The solid curve is a plot of $\sigma_0$, the even nucleon-nucleon cross section,  taken  from an analytically constrained global $\ln^2s$ fit which included the cosmic ray data.  The AGASA data are the triangles, the Fly's Eye point is the diamond and the HiRes point is the open diamond. 
}
}
\label{fig:hirespp}
\end{figure}
{\em Acknowledgments.}
 I wish to thank Peter Mazur , Larry Jones and the Aspen Center for Physics for making it possible for me to ``give'' this talk  {\em in absentia}.

\end{document}